\documentclass[reprint,amsmath,amssymb,aps,]{revtex4-1}

\usepackage{graphicx}
\usepackage{dcolumn}
\usepackage{bm}

\usepackage{wasysym} 
\usepackage{hyperref}
\usepackage{makecell}

\newcommand{\ket}[1]{\left| #1 \right>}

\begin{document}

\preprint{APS/123-QED}

\title{Long-term trapping of cold polar molecules}

\author{Dominik Haas$^*$, Claudio von Planta$^*$, Thomas Kierspel, Dongdong Zhang$^{\dagger,\ddagger}$ and Stefan Willitsch$^\dagger$}

\affiliation{Department of Chemistry, University of Basel, Klingelbergstrasse 80, 4056 Basel, Switzerland \\
$\dagger$ Corresponding authors. Electronic mail: dongdongzhang@jlu.edu.cn, stefan.willitsch@unibas.ch \\
$^*$ These authors contributed equally to the present work. \\
$\ddagger$ Present address: Institute of Atomic and Molecular Physics, Jilin University, Qianjin Avenue 2699, Changchun 130012, China
}

\date{\today}

\begin{abstract}
We demonstrate the long-term ($<$ 1 minute) trapping of Stark-decelerated OH radicals in their $X~^{2}\Pi _{3/2}~(\nu = 0,~J = 3/2,~M_{J} = 3/2,~f)$ state in a permanent magnetic trap. The trap environment was cryogenically cooled to a temperature of 17~K in order to efficiently suppress black-body-radiation-induced pumping of the molecules out of trappable quantum states and collisions with residual background gas molecules which usually limit the trap lifetimes. The cold molecules were kept confined on timescales approaching minutes, an improvement of up to two orders of magnitude compared to room-temperature experiments, at translational temperatures on the order of 25~mK. The present results pave the way for spectroscopic studies of trapped molecules with long interaction times enabling high precision, for investigations of cold collisions and reactions with very small reaction rates, for new avenues for the production of ultracold molecules via sympathetic cooling and for the realisation of new forms of hybrid matter with co-trapped atoms or ions. 

\end{abstract}

\maketitle

\section{Introduction}

Over the last two decades, impressive progress has been achieved in the generation of translationally cold molecules ~\cite{carr09a,vandemeerakker12a,willitsch12a,lemeshko13a,moses16a,willitsch17a} motivated by a range of promising applications~\cite{wall16a,borri16a,bohn17a,safronova18a}. These include spectroscopic precision measurements for the exploration of physics beyond the Standard Model~\cite{hudson06a,bethlem08a,bethlem09a,truppe13a,baron14a,borri16a,kozyryev17a,lim18a}, studies of collisions and chemical reactions at very low temperatures~\cite{willitsch08a,ospelkaus10b,hall12a,yan13a,chefdeville13a,vogels15a,wu17a,vogels18a}, and new approaches to quantum simulation~\cite{micheli06a,blackmore18a} and quantum information~\cite{demille02a,yelin06a}. Many of these experiments require the interaction of the cold molecules with either radiation or other particles in traps over extended periods of time. 

Several methods have been demonstrated for the trapping of cold molecules. Cold polar molecules produced by Stark deceleration~\cite{meerakker05a,quintero-perez13a} or by velocity selection followed by Sisyphus cooling have been stored in electrostatic traps~\cite{hoekstra07a,englert11a,seiler11a, zeppenfeld12a,prehn16a}. Paramagnetic molecules decelerated to low velocities~\cite{lu14a}, generated using a cryogenic-buffer gas source~\cite{weinstein98a} or produced by photodissociation \cite{eardley17a} have been confined in electromagnetic~\cite{weinstein98a,liu15a}, permanent magnetic \cite{sawyer07a,liu17a,akerman17a,eardley17a,mccarron18a,williams18} and very recently superconducting magnetic traps~\cite{segev19a}. Optical trapping of cold molecules formed by photoassociation~\cite{takekoshi98a} or magnetoassociation~\cite{ni08a} has also been demonstrated. Recently, the first magneto-optical traps for molecules have been implemented~\cite{hummon13a,barry14a}.

With few exceptions (see, e.g., \cite{tsikata10a,englert11a,zeppenfeld12a,prehn16a}), the trap lifetimes of cold polar molecules that have been achieved in previous studies range typically from milliseconds to a few seconds, which is far less than what has been achieved with neutral atoms in atom traps and particularly with ions in ion traps~\cite{wieman99a,willitsch12a}. There are several reasons for this. Energetic collisions with background-gas molecules, which can occur on timescales of seconds even under the ultrahigh-vacuum conditions of typical experiments, usually impart sufficient kinetic energy to the confined molecules in order to eject them from the shallow traps. Similarly, chemical reactions with background molecules remove density from the trap~\cite{janssen13a}. Moreover, in many experiments the lack of a continuous cooling mechanism for the trapped molecules limits their temperatures, densities and lifetimes. 
Of imminent importance for trapped polar molecules, however, is their interaction with the ambient black-body radiation (BBR) field which can continuously pump them into quantum states which are not magnetically or electrically trappable and therefore lead to the loss of their confinement~\cite{hoekstra07a}. 

These limits on the trap lifetimes represent a severe impediment for several applications of cold molecules. Precision spectroscopic measurements require long interaction times of the radiation with the molecules~\cite{carr09a,wall16a,borri16a}. Studies of chemical reactions with cross sections far below the collision limit, which represent the vast majority of all chemical processes, equally require long contact times between the collision partners to obtain a significant reaction yield. This problem is further aggravated by the low number densities of trapped molecules achieved so far. Consequently, studies of reactions with trapped molecules have thus far been limited to fast processes \cite{ospelkaus10b}. Similarly, many applications in quantum science also require long trapping and coherence times of the particles. Indeed, the capability to store and coherently manipulate cold ions for minutes has been one of the main reasons for the impressive success of ion-trap based approaches to quantum computing \cite{haeffner08a}.

\begin{figure*}
\includegraphics[width=0.95\textwidth]{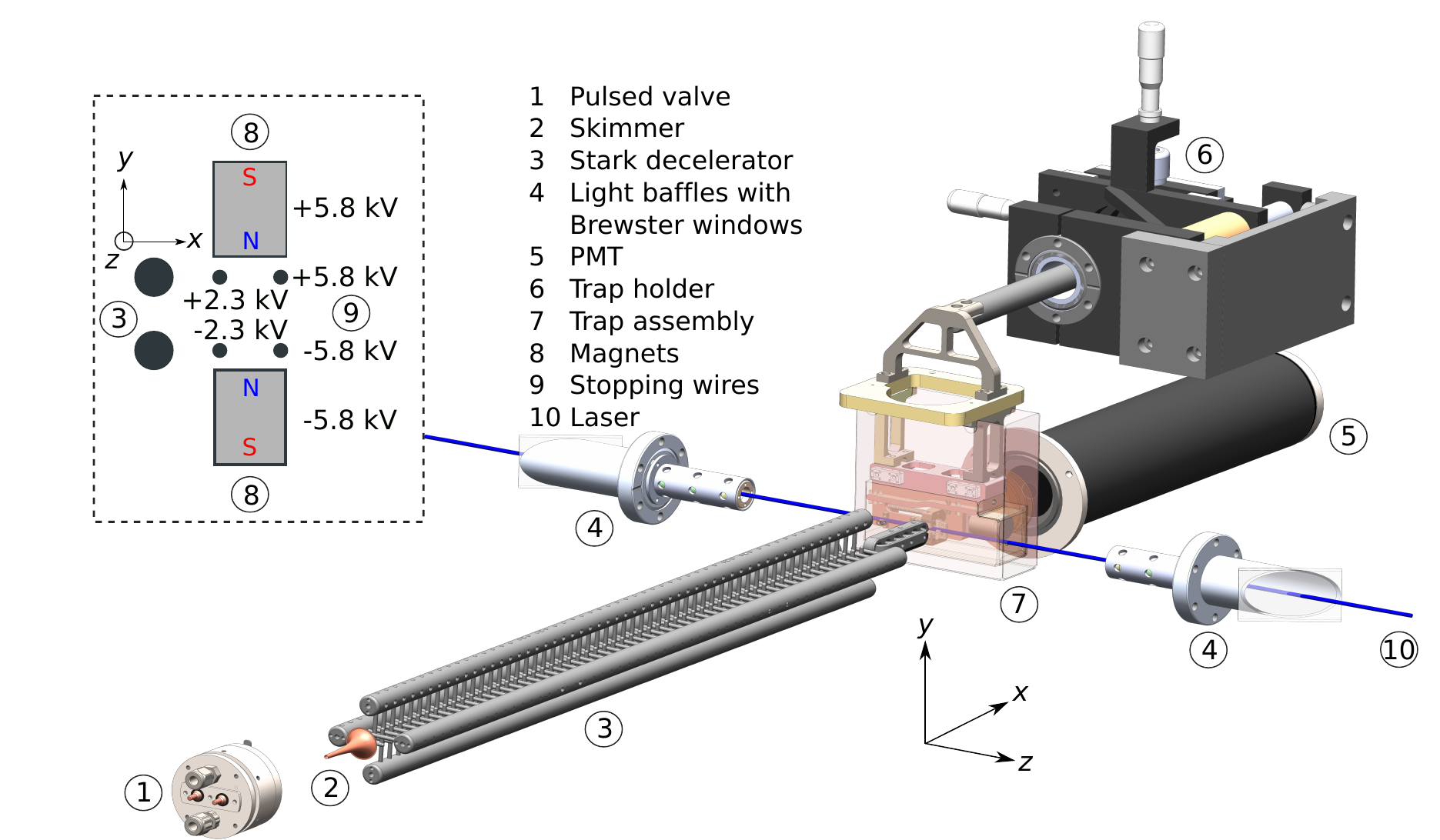}
    \caption{Experimental setup. Pulsed beams of internally cold OH radicals were generated by a pinhole discharge gas nozzle (1). After passing through a skimmer (2), the molecules were decelerated from velocities of 425~m/s down to 29~m/s by a 124-stage Stark decelerator (3). The translationally cold molecules were loaded into a permanent magnetic trap (7) by applying high voltages to the bar magnets (8) and stopping wires (9) acting as a last deceleration stage. The whole trap assembly was mounted on a $xyz$-translation stage (6) for fine adjustment of its position. Free-flying decelerated or trapped molecules were detected by laser-induced fluorescence (LIF) collected by a lens onto a photomultiplier tube (PMT) (5). The trap region was shielded from stray light using light baffles (4). The inset shows a  side-view schematic of the trapping region.} 
\label{fig:ht-expsetup} 
\end{figure*}

In this paper, we report the cryogenic magnetic trapping of Stark-decelerated molecules on timescales approaching minutes, exceeding previous room-temperature studies by one to two orders of magnitude. The cryogenic environment efficiently reduces the intensity of the ambient BBR field and further improves the vacuum conditions enabling a marked improvement of the trapping times. The trap lifetimes achieved here are similar to the ones achieved in previous cryogenic trapping experiments of molecules loaded from buffer-gas  \cite{tsikata10a} or velocity-selection sources \cite{englert11a,zeppenfeld12a, prehn16a}. The realisation of such long trapping times lays the foundations for applications of trapped cold molecules in precision spectroscopy, in studies of slow chemical processes at low energies and in the quantum technologies.

\section{Results and discussion}
\label{sec: results}

Our experimental setup is illustrated in Fig.~\ref{fig:ht-expsetup}. Packages of internally cold OH radicals were produced by an electric discharge of H$_2$O vapour seeded in 2.5 bar Kr gas during a supersonic expansion into high vacuum \cite{ploenes16a}. The molecule package propagated through a skimmer into a Stark decelerator \cite{haas17a} in which it was slowed down for subsequent loading into a cryogenic permanent magnetic trap (inset in Fig.~\ref{fig:ht-expsetup}). 
\subsection{Cryogenic magnetic trap.} The magnetic trap consisted of two Ni-coated bar magnets generating a quadrupolar magnetic field. We employed PrFeB as material for the magnets instead of the more widely used NdFeB because the latter exhibits a decreased magnetisation level at cryogenic temperatures due to a spin-reorientation transition~\cite{garcia00a}. By contrast, the remanence of the PrFeB magnets was specified to increase from 1.40~T at room temperature to 1.64~T under cryogenic conditions \cite{borgermann14a}. The entire trap assembly was mounted on a $xyz$-translation stage allowing fine adjustment of the position of the magnets with respect to the exit of the decelerator in order to optimise the loading efficiency. 
The trap was enclosed in a two-layer cryogenic shield consisting of copper plates cooled by a closed-cycle refrigerator. The cold head of the refrigerator was suspended from a spring-loaded assembly and connected to the cryogenic shield with copper braids in order to isolate the trap from vibrations. Temperatures of 17~K at the inner shield and of 53~K at the outer shield were thus obtained. As the BBR intensity scales with the temperature $T$ as $T^4$, the cryogenic environment effectively suppressed BBR pumping of the trapped molecules into untrapped states.  
The shielding of the trap from room temperature BBR was, however, not perfect because of apertures in the assembly necessary for admitting the molecular beam, for inserting laser beams enabling the spectroscopic probing of the trapped molecules and for collecting their laser-induced fluorescence.

\begin{figure}
\includegraphics[width=1\linewidth]{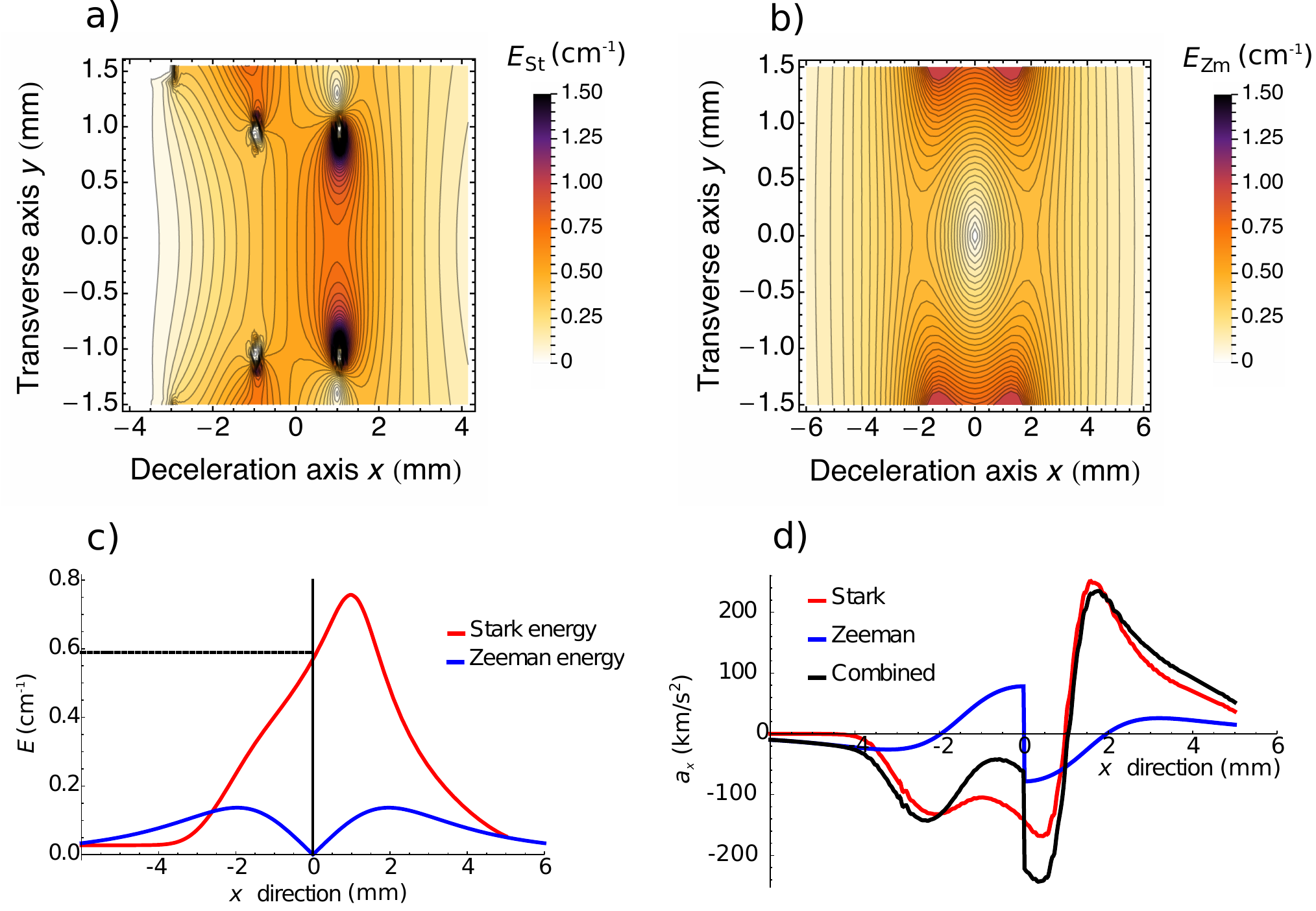}
    \caption{Stopping and trapping potentials. \textbf{a} Stark energies $E_{\text{St}}$ for OH in the $J = 3/2,~M_J = \pm 3/2,~f$ state resulting from applying an electric potential of $\pm 5.8$~kV on the bar magnets and the rear stopping wires. The stopping wires closer to the decelerator exit are kept at $\pm 2.3$~kV. \textbf{b} Zeeman energies $E_\text{Zm}$ for OH in the $J = 3/2,~M_J = + 3/2,~f$ state assuming a remanescence of the permanent magnets of 1.64~T. \textbf{c} Stark/Zeeman energy profile along the stopping direction ($y,z = 0$). The black vertical bar represents the trap centre and the horizontal dashed line corresponds to the kinetic energy of the synchronous molecule at the decelerator exit. \textbf{d} Longitudinal acceleration $a_x$ for OH molecules according to the Stark and Zeeman energies from \textbf{c}. All distances are given relative to the trap centre.}
\label{fig:ht-fields}
\end{figure}

\subsection{Stark deceleration and trap loading.} 
Packages of translationally cold OH radicals in the $X~v=0,J = 3/2, M_J = \pm 3/2, f$ state were produced by Stark deceleration using a 124-stage decelerator~\cite{haas17a}. Here, $v$ denotes the vibrational quantum number of the molecule, $J$ is the quantum number of its total angular momentum without nuclear spin and $M_J$ is the corresponding space-fixed projection quantum number. $f$ designates the parity label.

\begin{figure*}
\includegraphics[width=0.8\textwidth]{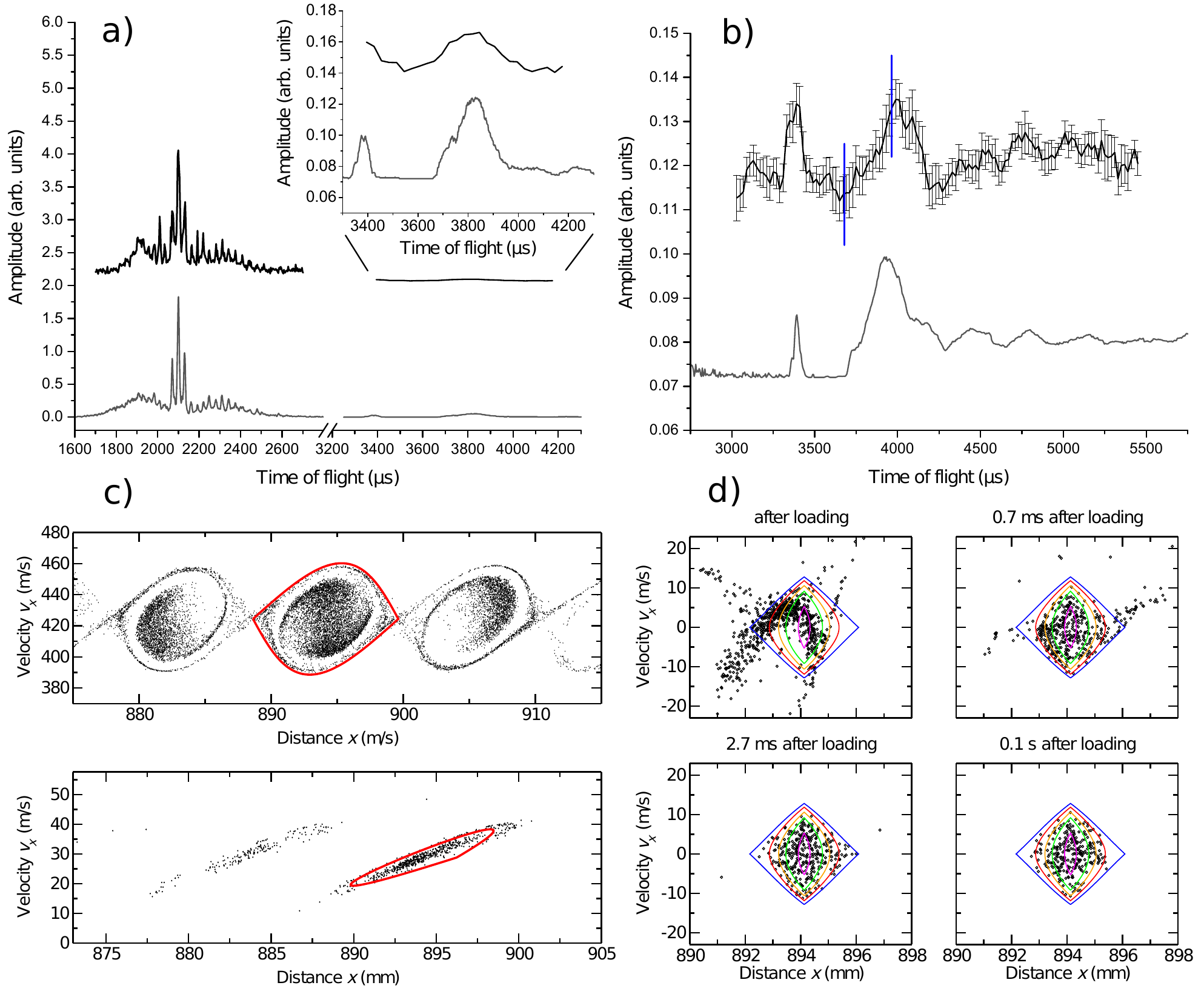}
    \caption{Time of flight (TOF) profiles of OH molecules and corresponding phase space distributions. All TOF profiles were recorded by collecting laser-induced fluorescence (LIF) 11.5~mm downstream from the exit of the Stark decelerator at the position of the trap centre. The distances are given with respect to the position of the valve. In \textbf{a} and \textbf{b}, the dark solid lines (upper traces) represent experimental results and the light solid lines (lower traces) Monte-Carlo trajectory simulations (see text). All traces are normalised with respect to the maximum signal level observed in the trace in panel \textbf{a} obtained in guiding mode ($v_\text{initial} = v_\text{final} = 425$~m/s). The inset in panel \textbf{a} displays a magnification of the TOF profile at the targeted loading velocity of $v_\text{target} = 29$~m/s. \textbf{b} TOF trace illustrating the final deceleration and the onset of trapping at room temperature. The period of application of the electric stopping field, by which the velocity of the molecule packet is reduced from $29$~m/s to $8$~m/s, is indicated by two blue vertical lines. \textbf{c} Top: Phase space distribution of the OH package at the trap centre for the decelerator operating in guiding mode (see panel \textbf{a}). Due to the large spatial and velocity spread of the molecular beam, a total of three phase-stable regions was loaded and transported through the decelerator. The packet indicated in red was selected for trap loading. Bottom: Phase space distribution at $v_\text{target} = 29$~m/s. The fastest phase-stable package is not shown. The phase-space volume is tilted due to the free flight from the decelerator to the trap. \textbf{d} Phase space evolution at different points in time after loading the trap, extracted from the simulations shown in \textbf{b}. The contour lines indicate the phase-space acceptance of the trap in increasing order of energy: 0.02 (magenta), 0.06 (green), 0.08 (yellow), 0.10 (red) and 0.12 (blue)~cm$^{-1}$. Uncertainties correspond to the standard error of 5 experimental repetitions in panel \textbf{b}.}
\label{fig:ht-exptcurves}
\end{figure*}

Efficient transfer of the decelerated OH package into the magnetic trap required a careful minimisation of particle losses during the loading procedure. Successful trap loading necessitates minimising the mismatch between the phase-stable volume transported by the decelerator and the trap. During free flight after exiting the decelerator, the molecule packages expand in both the longitudinal and transversal directions. Therefore, placing the trap close to the decelerator helps to prevent losses. In practice, this requirement is mitigated by the presence of the two cryogenic shields and a safety distance that has to be held towards the last high voltage electrode of the decelerator. As illustrated in the inset of Fig. \ref{fig:ht-expsetup}, the loading process was designed such that the bar magnets forming the magnetic trap also serve as a last electrostatic deceleration stage. This allows for loading the trap with molecules at higher velocities, since the final deceleration step occurs in close proximity to the trap centre \cite{sawyer07a,sawyer08a,stuhl12b}. In addition, four wire electrodes were spanned in between the magnets to introduce further degrees of freedom in shaping the stopping fields. 

The complexity of the stopping geometry and the considerable amount of parameters influencing the loading efficiency render it difficult to design and optimise the trapping process manually. Therefore, salient experimental parameters such as the voltages on the magnets and stopping wires as well as the phase angle of the deceleration process were numerically optimised using a mesh-adaptive search algorithm \cite{abramson19a,ledigabel11a}. For the experiments presented here, the potentials on the bar magnets were limited to $\pm 5.8$~kV by the electric breakdown strength of the trap. Fig. \ref{fig:ht-fields} \textbf{a} shows the stopping potentials expressed as Stark energies for OH in the $X~J = 3/2,~M_J = \pm 3/2,~f$ state allowing the removal of a maximum energy of 0.76~cm$^{-1}$ in the longitudinal direction along the centre line through the trap ($y,z = 0$). For comparison, the Zeeman trapping potential $E_\text{Zm}$ for the $X~J = 3/2,~M_J = + 3/2,~f$ state in between the two 1.64~T PrFeB magnets is depicted in Fig. \ref{fig:ht-fields} \textbf{b}. The resulting magnetic trap has a depth of 0.14~cm$^{-1}$ along the longitudinal $x$-direction (corresponding to a maximum velocity of 13.9~m/s for OH radicals in this state) as well as 0.33~cm$^{-1}$ (21.7~m/s) and 0.13~cm$^{-1}$ (13.3~m/s) along the transverse $y$- and $z$-directions, respectively. Note that from the packet of decelerated molecules, only those in the low field seeking $J = 3/2,~M_J = +3/2,~f$ Zeeman component can be confined magnetically \cite{sawyer07a,sawyer08a}. This comprises half of the molecules in the decelerated ensemble. As depicted in Fig. \ref{fig:ht-fields} \textbf{c}, the Stark energy slope extends beyond the trap centre and allows for matching the deceleration in Fig. \ref{fig:ht-fields} \textbf{d} to the velocity distribution of the incoming OH package.
 
 The OH molecules at various points in the assembly were pumped with a dye laser at a wavelength of around 282~nm and probed by monitoring their laser-induced fluorescence (LIF) at around 313~nm~\cite{ploenes16a}. The presence of the cryogenic shields, however, imposed a significant reduction in the solid angle under which fluorescence photons could be collected. In the present experiments, fluorescence from the molecules in the trap was acquired with a lens (\diameter = 6~mm, focal length $f = 6$~mm) mounted in the cryogenic shield. Therefore, inevitably the LIF signal levels obtained were low, but nonetheless sufficient for an unambiguous characterisation of the trap loading dynamics. Experimental time-of-flight (TOF) profiles of the molecules were validated against simulated TOF curves, as depicted in Fig. \ref{fig:ht-exptcurves} \textbf{a}. The simulations take into account contributions from both low-field-seeking components $M_J\Omega  = -9/4$ and $M_J\Omega  = -3/4$ which were transported through the decelerator ($\Omega$ denotes the quantum number of the projection of $\Vec{J}$ onto the molecular axis). As can be seen in Fig. \ref{fig:ht-exptcurves} \textbf{a}, the simulations accurately reproduce the experimental arrival time of the OH packages as well as the relative LIF signal intensities. All TOF traces in Fig. \ref{fig:ht-exptcurves} \textbf{a} were normalised to the signal in guiding mode (no deceleration, i.e., $v_\text{initial} = v_\text{final} = 425$~m/s). By comparing the experiments with the simulations, a spatial spread of the initial OH packet of 11.5~mm was deduced. Due to the large spatial and velocity spreads of the molecular beam, three phase-stable regions were loaded and transported through the decelerator as indicated by the phase-space diagrams depicted in Fig. \ref{fig:ht-exptcurves} \textbf{c}. This gives rise to a distinct triple-peak structure in the centre of the TOF profile of Fig. \ref{fig:ht-exptcurves} \textbf{a} of which the large middle peak is used for trap loading. Upon increasing the phase angle of the decelerator \cite{haas17a}, more energy is removed from the OH package per decelerator stage and the number of molecules (and therefore the signal level) decreases due to a reduction of the phase-stable volume. At a phase angle of $\phi = 55.468^\circ$, the target velocity for loading at $v_\text{target} = 29$~m/s was reached. The corresponding signal level was $\approx 70$ times lower than the one observed in guiding mode (see inset of Fig.  \ref{fig:ht-exptcurves} \textbf{a}). During the free flight over a distance of 11.5~mm between the decelerator exit and the detection point at the low final velocity of 29~m/s, the phase-space volume rotated significantly (see bottom panel of Fig. \ref{fig:ht-exptcurves} \textbf{c}) so that the signal in the TOF profile appears broadened. 

\subsection{Trapping of OH molecules at \\ room temperature and under cryogenic conditions.} 
To load the decelerated molecules into the magnetic trap, the stopping fields were switched on in between the points in time indicated by blue vertical lines in Fig. \ref{fig:ht-exptcurves} \textbf{b}. The TOF curve depicting the onset of trapping is shown in Fig. \ref{fig:ht-exptcurves} \textbf{b}. During the stopping process, the average velocity of the molecule packet was reduced from $29$~m/s to $8$~m/s for loading the trap. The TOF profile of the molecules right after trap loading displays an oscillatory pattern which is due to the molecule packet oscillating inside the trap. This behaviour is reproduced by the simulations. The decrease in signal intensity is attributed to the loss of molecules in magnetically high-field seeking states as well as the portion of molecules carrying sufficient kinetic energy to surmount the trap potential. The phase space evolution of the trapped OH cloud as extracted from the simulations is depicted in Fig. \ref{fig:ht-exptcurves} \textbf{d}. The simulations indicate that the trap loading is completed after 2.7~ms and after another 0.1~s the molecule package has obtained an average velocity of 6.0~m/s (corresponding to a kinetic energy of $E_k=0.026~$cm$^{-1}$ or $T=37$~mK).

\begin{figure}
\includegraphics[width=1\linewidth]{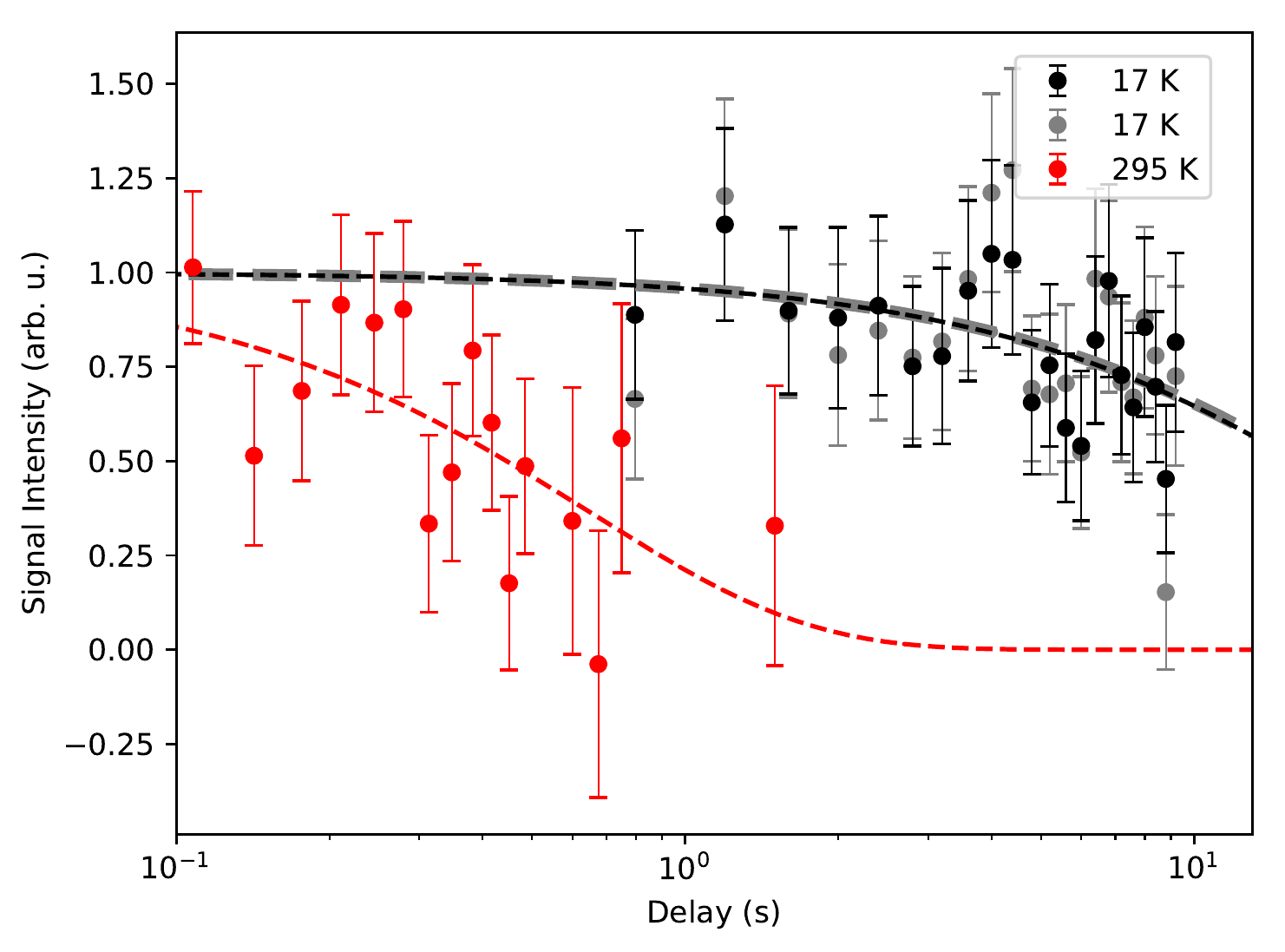}
    \caption{Lifetimes of trapped OH molecules. Fractional population of trapped OH radicals as a function of trapping time at room temperature (295~K, red) and under cryogenic conditions (17~K, black, grey). The grey  data points were obtained by subtracting experimentally determined background levels from each individual data point prior to fitting. The black data points result from the subtraction of a mean background value from the trapping signal. The dashed lines correspond to fits of the data to an exponential decay function. The 1/e lifetimes determined from the fits are $0.6(2)$~s at 295~K (red), $24(13)$~s (grey) as well as $23(8)$~s (black) at 17~K. The time origin coincides with switching off the stopping fields. Uncertainties correspond to the standard error of 500-1200 and 121 experimental cycles for the room-temperature and cryogenic data, respectively.}
\label{fig:trapping_life}
\end{figure}

\begin{table}
\caption{Calculated pumping rates out of the trapped $\ket{\nu = 0,~J = 3/2, F = 3/2, ~M_J = 3/2,~f}$ state as well as background collision rates and radiative lifetimes. The experimentally determined lifetimes are given in the last column. See text for details.}
\vspace{2mm}
\begin{tabular}{c|ccc|c}
        \makecell{Assemb.\\temp.} & \makecell{BG collision \\ rate (s$^{-1}$)} & \makecell{BBR pumping \\ rate (s$^{-1}$)} &  \makecell{BBR \\ LT (s)} & \makecell{exp. \\LT (s)} \\
        \hline
        \hline
        298~K & 1.2 & 0.40 & 2.8 & 0.6 (2) \\
        17~K & 0.026 & 0.030 & 81 & 23 (8)
    \end{tabular}
    \label{tab:pumprates}
\end{table}

The LIF signals of the trapped molecules as a function of the trapping time at room and cryogenic temperatures are shown in Fig.~\ref{fig:trapping_life}. At room temperature, the 1/e lifetime for the confinement of the molecules was determined to be $0.6(2)$~s. To assess whether the trap lifetimes are limited by BBR pumping or ejection of the trapped molecules by collisions with background gas, we modeled the trap lifetimes due to interaction with BBR and background gas collisions. For this purpose, we calculated the transition rates between the energy levels under the influence of BBR pumping and collisions with background gas molecules at a certain pressure and solved the relevant rate equations. We estimate a purely radiative lifetime of 2.8~s, which is in agreement with the results from Hoekstra \textit{et al.} \cite{hoekstra07a}. The fact that our experimental result is more than a factor of 4 lower suggests that in this regime the trap lifetimes are limited by collisions with background gas. The corresponding Kr gas pressure inside the trapping region was estimated to be $4\times10^{-8}$~mbar. This is a reasonable value considering that the fast part of the molecular beam is reflected off the surfaces of the cryogenic shield and the gas molecules remain in the trapping region for some time before they can escape through the apertures to be pumped away. 

Under cryogenic conditions (17~K), the lifetime increases to $23(8)$~s, a value about a factor of 3 smaller than the expected limit from BBR pumping (see Tab. \ref{tab:pumprates}). This result indicates that in this regime the lifetimes are considerably enhanced, but still limited by collisional processes. While most gases should efficiently freeze out on the cryogenic shields surrounding the trap, these residual collisions could originate from gas particles streaming into the trapping region through the apertures in the cryogenic shield and hitting the OH molecules before they freeze out on the surfaces. Another likely contribution are collisions with hydrogen molecules which are not expected to freeze out efficiently on the surfaces under the UHV conditions and cryogenic temperatures of the present experiment. Assuming that the trap lifetime is limited by the collision rate with H$_2$ molecules at 17~K, an effective H$_2$ pressure of $3 \times 10^{-11}$~mbar in the trapping region can be deduced from this lifetime. This value is on the same order of magnitude as the typical pressures measured by a pressure gauge in the trap chamber surrounding the cryogenic shield. 

\section{Conclusion}
\vspace{-1mm}
The observed enhancement of the trapping time of the cold OH molecules in the cryogenic environment amounts to a factor of 40 with respect to room temperature and to almost a factor of 10 compared to the previous result of Ref. \cite{hoekstra07a}. The current trap lifetimes are attributed to collisions with residual gas. This conclusion is supported by the observation that the measured trap lifetimes fluctuated slightly on timescales of several days probably as a result of slightly changing vacuum conditions in the setup. The effective trap lifetimes, therefore, critically depend on the experimental conditions and the values presented here are typical values achievable with the present setup.\\
Consequently, a further increase of the trap lifetime should be obtainable by transporting the trapped molecules into a "darker" region of the assembly with an even further reduced exposure to gas as well as room-temperature BBR entering the trapping region from outside the cryogenic shield. This facility is currently being implemented in our setup.\\
The trapping times achieved in the present study are sufficiently long to enable prolonged spectroscopic measurements on the trapped molecules and the measurement of slow reactive processes, enabling a new range of applications for cold trapped molecules. In this context, an intriguing perspective is the simultaneous trapping of the cold molecules with cold trapped ions \cite{willitsch15a, willitsch17a} which paves the way for ion-neutral hybrid systems of purely molecular matter. Experiments in this direction are currently underway in our laboratory.

\begin{acknowledgments}
D.Z. acknowledges the financial support from Freiwillige Akademische Gesellschaft (FAG) Basel and the Research Fund for Junior Researchers of the University of Basel. We thank Prof. Sebastiaan Y.T. van de Meerakker (Radboud University) for helpful discussions. This project is funded by the Swiss National Sciene Foundation, grant nr. 200020\_175533, and the University of Basel. We would like to thank P. Kn{\"o}pfel, G. Martin and G. Holderied from the departmental workshops for technical support.
\end{acknowledgments}

\vspace{2mm}
\section*{Author contributions}
D.H. and C.P. contributed equally to the presented work. D.H., C.P., T.K. and D.Z. conducted the experiments and analysed the data. S. W. conceived and D.Z. and S.W. supervised the project. D.H. designed and optimised the trap loading process based on trajectory simulations. C.P. calculated the black-body radiation and background collision limited trapping lifetimes and implemented the cryogenic trap. All authors participated in writing the manuscript. 

\bibliography{finref}

\end{document}